\documentstyle[preprint,floats,prd,aps]{revtex}
\begin{document}
\preprint{\vbox {\hbox{RIKEN-AF-NP-279}}}
%\preprint{\vbox {\hbox{OCHA-PP-105}}}

\draft
\title{Nonperturbative QCD contribution to gluon-gluon-$\eta'$ vertex} 
\author{Mohammad R. Ahmady$^a$\footnote{Email: ahmady@riken.go.jp}, Victor 
Elias$^b$\footnote{Email:zohar@apmaths.uwo.ca} and Emi 
Kou$^c$\footnote{Email: kou@fs.cc.ocha.ac.jp}}
\address{
$^a$LINAC Laboratory, The Institute of Physical
and Chemical Research (RIKEN)\\ 2-1 Hirosawa, Wako, Saitama 351-01,
Japan \\
$^b$Department of Applied Mathematics, University of Western Ontario\\
London, Ontario, Canada N6A 5B7\\
$^c$ Department of Physics, Ochanomizu University \\
1-1 Otsuka 2, Bunkyo-ku,Tokyo 112, Japan}
\date{January 1998}
\maketitle
\begin{abstract}
We investigate the momentum dependence of the $g^*-g-\eta'$ vertex form factor in the presence of the nonperturbative QCD vacuum condensate.  The behaviour of this form factor is crucial for,  among other things, some proposed mechanisms devised to explain the large branching ratio for fast $\eta'$ production in charmless B decays observed by CLEO.  We show that the leading momentum dependence of this vertex form factor is not altered when the contribution of nonperturbative quark condensate is considered.
\end{abstract}
%
%\pacs{13.20.He, 12.20.Ds}
\newpage
The recent observation at CLEO of a large branching ratio for $B\to X_s\eta'$ decay\cite{cleo} has led to various attempts for a theoretical explanation.  
Most of these attempts rely on $\eta'$ being  produced via its anomalous coupling to two gluons, which is parametrized by the form factor $H(q^2,p^2,m_\eta^2)$ with $q$ and $p$ being the four momenta of the gluons \cite{as,ht,aks}.  The momentum behaviour of this form factor is crucial for the success or failure of these proposed mechanisms.  For example, $H(q^2,0,m_{\eta'}^2) \approx H(0,0,m_{\eta'}^2)$ has been assumed in Ref [2], so as to obtain results which agree with experiment.  It has been pointed out, however, that a sigma-model calculation of $g^* - g - \eta'$ vertex ($g^*$ and $g$ are off-shell and real gluons, respectively) leads to a significant momentum dependence of the form $H(q^2,0,m_{\eta'}^2) \sim 1 / (q^2 - m_{\eta'}^2)$ \cite{kp}.  In attempting to account for the $B \to X_s \eta'$ branching ratio, it is of interest to determine whether this leading momentum dependence is altered by nonperturbative QCD contributions. In this paper, we calculate the quark condensate component of the nonperturbative contribution to $H$ to investigate whether the insertion of the $< \bar ff>$ augmented quark propagator in the triangle loop alters the leading momentum dependence of the vertex form factor.
The purely-perturbative form factor $H(q^2,0,m_{\eta'}^2)$ for the 
vertex $g^*-g-\eta'$ (Fig. 1) is found to be
\begin{equation}
V_{\mu\nu}=\Sigma_{f=u,d,s}\frac{g^{\eta'}_fg_s^2m_f}{2\pi^2}
H(q^2,0,m_{\eta'}^2)\epsilon_{\mu\nu\alpha\beta}q^\alpha p^\beta\;\; ,
\end{equation}
with
\begin{eqnarray}
H(q^2,0,m_{\eta'}^2)&=&\displaystyle\frac{1}{q^2-m_{\eta'}^2} \left [I(\frac{m_{\eta'}^2}{m_f^2})-I(\frac{q^2}{m_f^2})\right ]\;\; ,\\ \nonumber I(x^2)&=&\displaystyle\left \{^{\displaystyle -2{\rm arcsin}^2 \frac{\vert x\vert}{2}\;\;\; 0\le x^2\le 4\;\; ,}_{\displaystyle{ 2 \left [{\rm Ln}(\frac{\vert x\vert}{2}+\sqrt{\frac{x^2}{4}-1})-\frac{i\pi}{2}\right ]}^2\;\; x^2\ge 4\;\; .}\right.
\end{eqnarray}
$g^{\eta'}_f$ is the coupling of $\eta'$ to the quark of flavour $f$, and $g_s$ is the strong interaction coupling constant.  We note from Eq. (2) that the leading momentum dependence of $H\sim 1/(q^2-m_{\eta'}^2)$ indicates a significant suppression of this form factor at large $q^2$.  As a result, the mechanism suggested in Ref. [2] for $B \rightarrow X_s \eta'$ decay falls short of experimental data by at least an order of magnitude. A reconciliation of this mechamism to experiment, however, may be possible from a hardening of the leading momentum dependence of $H$ by the nonperturbative content of QCD, content which is not taken into account in Eq. (2). We investigate this possibility by calculating the quark-condensate contributions to the form factor $H$.  QCD condensates characterize the nonperturbative content of the QCD vacuum, and therefore, the inclusion of a quark-condensate insertion should give us some indication of the leading nonperturbative contributions to $H$. 
We proceed first by replacing the usual perturbative quark propagator in Fig. 1 with the full quark propagator $S(k)$ \cite{baes}: 
\begin{equation}
S(k)=S^{\rm P}(k)+S^{\rm NP}(k)=-i\int d^4 xe^{ik.x}<\Omega\vert T\Psi 
(x)\bar\Psi (0)\vert\Omega >\;\; ,
\end{equation}
where $\vert\Omega >$ is the nonperturbative QCD vacuum.  As a result, the nonperturbative contributions to the quark propagator can be written as \begin{equation} iS^{\rm NP}(k)=\int d^4xe^{ik.x}<\Omega\vert :\Psi (x)\bar\Psi (0):\vert\Omega >\;\; .
\end{equation}
The nonlocal vacuum expectation value (vev) in Eq. (4) can be expanded in 
terms of local condensates \cite{vss}.  To ascertain how leading 
nonperturbative QCD contributions affect the form factor $H$, only
the lowest-dimensional (quark-condensate) component of the expansion
is considered below.  The quark-condensate projection of the
nonperturbative quark propagator is taken from Ref. [6],
\begin{equation}
iS^{\rm NP}(k)={(2\pi )}^4(\gamma_\mu k^\mu +m_f)F(k),
\end{equation}
where the Fourier transform of $F(k)$ is given by
\begin{equation}
\int d^4ke^{-ik.x}F(k)=-\frac{<\bar ff>}{6m_f^2}
\frac{J_1(m_f\sqrt{x^2})}{\sqrt{x^2}}\;\; .
\end{equation}
$F(k)$ has the following important on-shell property \cite{baes}:
\begin{equation}
(k^2-m_f^2)F(k)=0\;\; .
\end{equation}
The dimension-3 quark condensate $<\bar ff>$ is the vev of the normal ordered local two-quark fields, i.e.
\begin{equation}
<\bar ff>=<\Omega\vert :\bar\Psi (0)\Psi (0):\vert\Omega >\;\; .
\end{equation}
Using the Feynman rule of the Eq. (5), one can obtain the nonperturbative $<\bar ff>$ contribution to $H$ in a straightforward manner.  The relevant Feynman diagrams are illustrated in Fig. 2 where the nonperturbative quark propagator $S^{\rm NP}$ is depicted by a disconnected line with two dots.  For example, the expression for Fig 2a is written as follows:
\begin{eqnarray}
\nonumber ^{(2a)}V_{\mu\nu}^{<\bar ff>}&=&\displaystyle -4 g_f^{\eta'}
g_s^2m_f\epsilon_{\mu\nu\alpha\beta} \; q^\alpha p^\beta \\
&\times &\displaystyle\int d^4k\int_0^1 dz\frac{F(k)}{{\left [-zq^2-2k
\cdot \{ z(q+p)-p\}\right ]}^2}\;\; ,
\end{eqnarray}
where the gluon with four-momentum $p$ is taken to be on-shell.  To 
proceed, 
we use the following formulae:
\begin{eqnarray}
\nonumber \displaystyle\int d^4k\frac{F(k)}{{\left [a-k \cdot 
v+i\epsilon\right ]}^2}
&=& \displaystyle - \int d^4kF(k)\int^\infty_0d\eta\eta 
e^{i\eta (a-k \cdot v+i\epsilon )}\;\;,\\
&=&\nonumber\displaystyle \frac{<\bar ff>}{6m^2_f}\int^\infty_0d\eta 
e^{i\eta a}
\frac{J_1(m_f\eta\sqrt{v^2})}{\sqrt{v^2}}\;\; ,\\
&=& \displaystyle \frac{<\bar ff>}{6m_f^3v^2}
\left (1-\frac{1}{\sqrt{1-\frac{m_f^2v^2}{a^2}}}\right )\;\; ,
\end{eqnarray}
where the second line is obtained from Eq. (6), and the final line
from differentiation (with respect to $a$) of a tabulated integral
\cite{gr}, corresponding to the second line of (10).  Consequently, the 
contribution of Fig 2a (Eq. (9)) is
\begin{eqnarray}
\nonumber
^{(2a)}V_{\mu\nu}^{<\bar ff>}&=&\displaystyle -\frac{g_f^{\eta'}g_s^2<\bar 
ff>}{6m_f^2}\epsilon_{\mu\nu\alpha\beta}q^\alpha p^\beta \\
&\times&\displaystyle\int^1_0dz\frac{1}{z(zm_{\eta'}^2+q^2-m_{\eta'}^2)}
\left [1-\frac{1}{\sqrt{1-\frac{4m_f^2(zm_{\eta'}^2+q^2-
m_{\eta'}^2)}{zq^4}}}\right
 ]\;\; .
\end{eqnarray}
Expressions for Figs. 2b and 2c are calculated in the same way, leading to the following aggregate quark-condensate contribution to $g^*-g-\eta'$ vertex:
\begin{equation}
\displaystyle V_{\mu\nu}^{<\bar ff>}=\Sigma_{f=u,d,s} \frac{g_f^{\eta'}g_s^2m_f}{2\pi^2}H^{<\bar ff>}(q^2,0,m_{\eta'}^2)
\epsilon_{\mu\nu\alpha\beta}q^\alpha p^\beta\;\; ,
\end{equation}
with
\begin{eqnarray}
\displaystyle H^{<\bar ff>}(q^2,0,m_{\eta'}^2)= 
-\frac{\pi^2<\bar ff>}{3m_f^3}\frac{1}{q^2-
m_{\eta'}^2}G(\frac{q^2}{m_{\eta'}^2},
\frac{m_f^2}{m^2_{\eta'}})\;\; ,
\end{eqnarray}
where
\begin{eqnarray}
\nonumber\displaystyle G(x,y)= {\rm Ln}\left [\frac{4y^3\left [(x-1)(\sqrt{x(x-4y)}+2y)-x^2+x{(x-1)}^2\right]} {\left [2y(x-1)+x^3-2x^2\right ]\left [(2y-1)\sqrt{1-4y}+2y^2-4y+1\right ] \left [\sqrt{x(x-4y)}+x-2y\right ]}\right ]\;\; .
\end{eqnarray}
We note that the leading momentum dependence of the quark condensate contribution to the vertex form factor is indeed the same as in (2);  {\it i.e.}, $H^{<\bar ff>}\sim 1/(q^2-m^2_{\eta'})$.
Therefore, we conclude that the lowest-dimensional contribution from nonperturbative QCD {\it does not} alter the significant suppression of the $g^* - g - \eta'$ vertex for large gluon virtualities.  Whether or not this result is valid when higher dimensional condensates are considered remains to be investigated.  However, one would anticipate the effect of higher-dimensional order parameters to be negligible at large momenta, suggesting that the momentum-dependence of $g^* - g - \eta'$ vertex form factor is not altered appreciably by nonperturbative QCD effects.
\section*{acknowledgements}
M. A. acknowledges support from the Science and Technology Agency of Japan under an STA fellowship.  V.E. acknowledges support from the Natural Sciences and Engineering  Research Council of Canada.

\newpage
{\center \bf \huge Figure Captions}
\vskip 3.0cm
\noindent
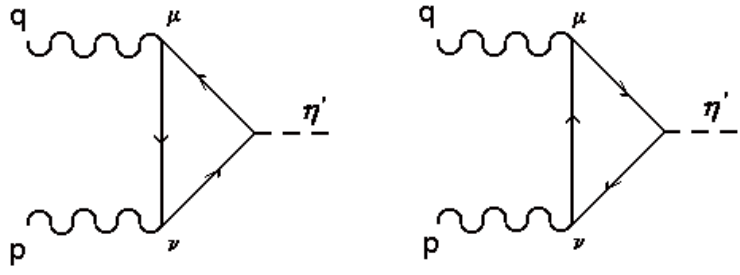
{\bf Figure 1}: Quark loop triangle diagram for $g-g-\eta'$ vertex. \\
\vskip 0.5cm
\noindent
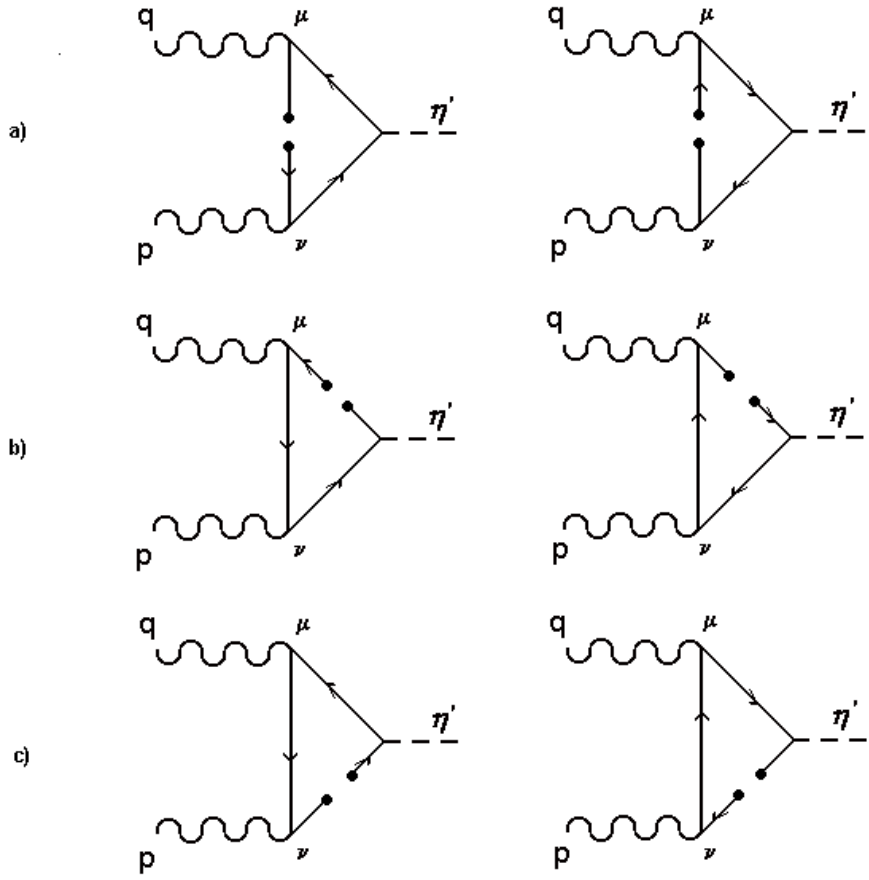
{\bf Figure 2}: Nonperturbative quark condensate contributions to $g-g-
\eta'$ 
vertex. \\
\vskip 0.5cm

\end{document}